\documentclass[a4paper, 10 pt, conference]{IEEEtran} 
\usepackage[singlespacing]{setspace} 
\usepackage{cite}
\usepackage{amsmath,amssymb,amsfonts}
\usepackage{graphicx}
\usepackage{textcomp}
\usepackage{xcolor}
\usepackage{cite}
\usepackage{physics}
\usepackage{amsmath}
\usepackage[utf8]{inputenc}
\usepackage{graphicx}
\usepackage{subcaption}
\usepackage{tikz}
\usetikzlibrary{quantikz}
\usepackage{tabularx}
\usetikzlibrary{arrows}
\usepackage{adjustbox}
\usepackage{algorithm}
\usepackage{algpseudocode}
\usepackage{balance}

\usepackage[T1]{fontenc}
\def\BibTeX{{\rm B\kern-.05em{\sc i\kern-.025em b}\kern-.08em
    T\kern-.1667em\lower.7ex\hbox{E}\kern-.125emX}}
\def \sys {\textit{QHFP}}

\def\BibTeX{{\rm B\kern-.05em{\sc i\kern-.025em b}\kern-.08em
    T\kern-.1667em\lower.7ex\hbox{E}\kern-.125emX}}
\begin{document}

\title{Device-independent Quantum Fingerprinting for Large Scale Localization}

\author{\IEEEauthorblockN{Ahmed Shokry}
	\IEEEauthorblockA{\textit{Dept. of Computer Science and Engineering} \\
		\textit{American University in Cairo}\\
		Cairo, Egypt \\
		ahmed.shokry@aucegypt.edu}
	\and
	\IEEEauthorblockN{Moustafa Youssef}
	\IEEEauthorblockA{\textit{Dept. of Computer Science and Engineering} \\
		\textit{American University in Cairo \& Alexandria University}\\ 
	Cairo, Egypt\\
	moustafa-youssef@aucegypt.edu} 
}

\maketitle

\begin{abstract}
Although RF fingerprinting is one of the most commonly used techniques for localization, deploying it in a ubiquitous manner requires addressing the challenge of supporting a large number of heterogeneous devices and their variations. 
 We present \sys{}, a device-independent quantum fingerprint matching algorithm that addresses two of the issues for realizing worldwide ubiquitous large-scale location tracking systems: storage space and running time as well as devices heterogeneity. In particular, we present a quantum algorithm with a complexity that is exponentially better than the classical techniques, both in space and running time. \sys{} also has provisions for handling the inherent localization error due to building the large-scale fingerprint using heterogeneous devices. We give the details of the entire system starting from extracting device-independent features from the raw RSS, mapping the classical feature vectors to their quantum counterparts, and showing a quantum cosine similarity algorithm for fingerprint matching.

We have implemented our quantum algorithm and deployed it in a real testbed using the IBM Quantum machine simulator. Results confirm the ability of \sys{} to obtain the correct estimated location with an exponential improvement in space and running time compared to the traditional classical counterparts. In addition, the proposed device-independent features lead to more than 20\% better accuracy in median error.   This highlights the promise of our algorithm for future ubiquitous large-scale worldwide device-independent fingerprinting localization systems. 
\end{abstract}


\section{Introduction}
\label{sec:introduction}


Recent years have witnessed the advent of the field of quantum computing (QC), with a variety of large corporations and startups investing in quantum computing \cite{knight2017ibm, kelly2018preview, tannu2018case}. QC algorithms can solve problems that are infeasible to be solved by classical computers \cite{qbible}, providing exponential gains in some cases \cite{shor1999polynomial}. This opens the door for investigating quantum algorithms advantages in new fields such as worldwide location tracking. 

%

Large-scale worldwide fingerprinting localization are commonly used indoor \cite{elhamshary2017towards, youssef2015towards, aly2013new, elbakly2016robust} and outdoor \cite{shokry2018deeploc, shokry2020dynamicslam, gu2021effect, aly2015lanequest, aly2015semmatch} due to their accuracy. The fingerprinting techniques work in two phases: the offline RF fingerprint building phase and the online tracking phase. In the offline phase, the received signal strength (RSS) coming from the different reference points (RPs\footnote{These reference points can be, e.g., WiFi access points; cellular cell towers; or Bluetooth beacons.}) in the environment are recorded at the different discrete locations. In the online phase, the RSS coming from RPs at unknown locations are matched aganist the fingerprint and the closest location in the RSS space becomes the estimated location. However, when the training devices that are used to build the fingerprint and the test devices that are used for localization during the online phase are different, the system accuracy severely degrades. Hence, deploying such systems on different phone types is not a straightforward task as the RSS readings vary for the different kinds of phones, even at the same location and time. This problem, which is known as the device heterogeneity problem \cite{ibrahim2013enabling, ibrahim2010cellsense, ibrahim2011cellsense, park2011implications,shokry2017tale}, can prevent ubiquitous large-scale  fingerprinting-based localization.
Moreover, all the traditional fingerprinting techniques need to match the online RSS measurements from the different RPs to the RSS measurements at each fingerpirnt location, making their time and space complexity $o(MN)$, where $M$ is the number of fingerprint locations and $N$ is the number of RPs. This can hinder deploying the traditional outdoor/indoor localization in a the large-scale worldwide settings, especially for IoT environments, where the number of RPs in an environment can be significant. 

In this paper, we present \sys{}: a \textbf{\textit{practical}} device-independent quantum fingerprint-based location tracking algorithm that requires space and runs in $o(M \log(N))$. 
 We show the details of how to extract device-independent features from the received signal strength measurements, how to  construct the quantum fingerprint, how to encode the extracted features in quantum states, and finally; how to calculate the quantum similarity between the online features and the offline ones stored in the fingerprint. 

\addtolength{\topmargin}{+0.1cm}
We validate our quantum algorithm in a real testbed. In addition, we quantify its performance using simulations on an IBM quantum computer. The results show the ability of \sys{} to correctly obtain the estimated location with the same accuracy as its classical counterparts. This comes with an \textbf{\textit{exponential}} enhancement compared to the traditional classical fingerprinting techniques in \textbf{\textit{both space and time}}. In addition, \sys{} provides 20\% enhancement in  median localization error in addressing the heterogeneity problem. 

The remaining sections are organized as follow: we begin with a brief background on quantum computing in Section~\ref{sec:background}. Section~\ref{sec:qfp} provides the details of our device-independent quantum fingerprint localization algorithm. Details of the implementation and evaluation results are presented and discussed in Sections~\ref{sec:eval}. Finally, Section~\ref{sec:conclude} concludes the paper.

\section{Background on Quantum Computing}
\label{sec:background}

%

\newcommand\filla{rgb:black,1;white,7}
\newcommand\fillb{teal}
\newcommand\fillc{rgb:black,1;white,7}
\newcommand\filld{white!60!blue}

\newcommand*{\gateStyle}[1]{{\textsf{\bfseries #1}}}
\newcommand*{\hGate}{\gateStyle{H}}
\newcommand*{\xGate}{\gateStyle{X}}
\newcommand*{\circuitH}{\gate[style={fill=\filla},label style=black]{\textbf{\hGate{}}}}
\newcommand*{\circuitX}{\gate[style={fill=\fillb},label style=white]{\textnormal{\xGate}}}


In this section, we give a brief background on the basic concepts of quantum computing that we will build on in the rest of the paper~\cite{shokry2021quantum, shokry2021towards}.

A quantum bit (qubit) is the basic unit of information and is analogue to the classical bit. Contrary to classical bits, a qubit can exist in a \textbf{superposition} of the zero and one states.  This superposition is what allows quantum computations to work on both states at the same time. This is often referred to as quantum parallelism. Qubits can have various physical implementations, e.g. the polarization of photons. 

Formally, the Dirac notation is commonly used to describe the state of a qubit as $\ket{\psi} = \alpha \ket{0} + \beta \ket{1}$, where $\alpha$ and $\beta$ are complex numbers called the amplitudes of classical states $\ket{0}$ and $\ket{1}$, respectively. The state of the qubit is normalized, i.e. $\alpha^2 + \beta^2 =1$. When
the state $\ket{\psi}$ is measured, only one of  $\ket{0}$ or $\ket{1}$ is observed, with probability  $\alpha^2$ and $\beta^2$, respectively. The measurement process is destructive, in the sense that the state collapses to the value $\ket{0}$ or $\ket{1}$ that has been observed, losing the original amplitudes  $\alpha$ and $\beta$ \cite{qbible}.

Operations on qubits are usually represented by gates, similar to a classical circuit.
 An example of a common quantum gate is the NOT gate (also called Pauli-X gate) that is analogous to the not gate in classical circuits. In particular, when we apply the NOT gate to the state $\ket{\psi_0} = \alpha \ket{0} + \beta \ket{1}$, we get the state $\ket{\psi_1} =  \beta \ket{0} + \alpha \ket{1}$. Gates are usually represented by unitary matrices 
while states are represented by column vectors\footnote{The ket notation $\ket{.}$ is used for column vectors while the bra notation $\bra{.}$ is used for row vectors.}. The matrix for the NOT gate is
$\begin{bmatrix}
	0 & 1\\
	1 & 0
\end{bmatrix}
$ and the above operation can be written as $\ket{\psi_1} = NOT(\ket{\psi_0}) = \begin{bmatrix}
	0 & 1\\
	1 & 0
\end{bmatrix} \begin{bmatrix}
\alpha\\
\beta
\end{bmatrix}.
$

\begin{figure}[!t]
	\centering
	{\includegraphics[width=0.2\textwidth]{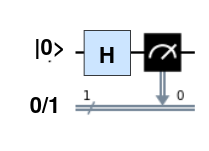}}
	\caption{An example of quantum circuit. Single lines represent quantum information and double lines represent classical information.}
	\label{fig:simplecircuit}
\end{figure}

Another important gate is the Walsh–Hadamard gate, $H$, that maps $\ket{0}$ to $\frac{1}{\sqrt{2}} (\ket{0}+ \ket{1})$, i.e. a superposition state with equal probability for $\ket{0}$ and $\ket{1}$; and maps $\ket{1}$ to $\frac{1}{\sqrt{2}} (\ket{0}- \ket{1})$. Figure~\ref{fig:simplecircuit} shows a simple quantum circuit. Single lines carry quantum information while double lines carry classical information (typically after measurement). The simple circuit applies an $H$ gate to state $\ket{0}$, which produces the state $\frac{1}{\sqrt{2}} (\ket{0}+ \ket{1})$ at the output of the gate. The measurement step produces either 0 or 1 with equal probability (the squared amplitude of the measured state). The state collapses to the observed classical bit value. 

It is important to note that the concept of quantum \textbf{interference} is at the core of quantum computing. Using quantum interference, one uses gates to cleverly and intentionally bias the content of the qubits towards the needed state, hence achieving a specific computation result. 

The notion of qubit can be extended to higher dimensions using a quantum register.
A quantum register $\ket{\psi}$, consisting of $n$ qubits, lives in a $2^n$-dimensional
complex Hilbert space $\mathcal{H}$. Register $\ket{\psi} = \sum_{0}^{2^n-1} \alpha_i \ket{i}$ is specified by complex numbers $\alpha_0, ..., \alpha_{2^n-1} $, where  $\sum |\alpha_i|^2 =1$. Basis state $\ket{i}$ denotes the binary encoding of integer $i$. We use the tensor product $\otimes$ to compose two quantum systems. For example,
we can compose the two quantum states  $\ket{\psi} = \alpha \ket{0} + \beta \ket{1}$ and  $\ket{\phi} = \gamma \ket{0} + \delta \ket{1}$ as $\ket{\omega} = \ket{\psi} \otimes \ket{\phi}= \alpha \gamma \ket{00} + \alpha \delta \ket{01}+ \beta \gamma \ket{10}+ \beta \delta \ket{11}$. %

\addtolength{\topmargin}{+0.1cm}

\begin{figure}[!t]
	\begin{subfigure}{0.5\textwidth}
		\centering
		{\includegraphics[width=0.35\textwidth]{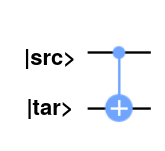}}
		\caption{The CNOT gate. }
		\label{fig:cnot}
	\end{subfigure}%
	
	\begin{subfigure}{0.5\textwidth}
		\centering
		{\includegraphics[width=0.35\textwidth]{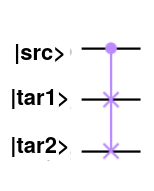}}
		\caption{The CSWAP gate.}
		\label{fig:cswap}
	\end{subfigure}
	
	\caption{Simple controlled gates. (a) The NOT gate is applied to the target qubit, if and only if, the source qubit is $\ket{1}$. (b) The two target qubits are swapped, if and only if, the source qubit is $\ket{1}$.}\label{fig:cntr_gate}
\end{figure}

Controlled gates act on multiple qubits, where one or more qubits act as a control for some operation on the other qubits (Figure~\ref{fig:cntr_gate}). Figure~\ref{fig:cnot} illustrates the controlled NOT gate (CNOT). When the source qubit is $\ket{1}$, the NOT operation will be applied to the target qubit. Figure~\ref{fig:cswap} further shows the controlled SWAP gate (CSWAP) with three qubits as input. In the CSWAP gate, the swap operation is performed on the target wires, if and only if, the source line is $\ket{1}$. This can be used to ``\textbf{entangle}'' qubits together. Entangled qubits are correlated with one another, in the sense that information on one qubit will reveal information about the other unknown qubit, even if they are separated by large distance \cite{qbible}.

A common way to describe a quantum algorithm is to use a quantum circuit, which is a combination of the quantum gates (e.g as in Figure~\ref{fig:simplecircuit}). The input to the circuit is a number of qubits (in quantum registers) and the gates act on them to change the combined circuit state using superposition, entanglement, and interference to reach a desired output state that is a function of the algorithm output. The final step is to measure the output state(s),  which reveals the required information.


\section{QHFP: A Device-independent Quantum Fingerprinting Localization System}
\label{sec:qfp}
In this section, we present the detailed description of our \sys{} quantum device-independent fingerprinting localization algorithm. 

The fingerprinting techniques work in two phases: the offline RF fingerprint building phase and the online tracking phase. In the offline phase, the RSS coming from the different RPs in the environment are recorded at the different discrete locations. In the online phase, the RSS coming from RPs at unknown locations are matched aganist the fingerprint and the closest location in the RSS space becomes the estimated location. 
Hence, fingerprint localization can capture the relation between RSS coming from the different RPs in the environment and user location \cite{youssef2005horus, youssef2006location, mohssen2014s}.

There are different metrics that can be used to match the online RSS with the fingerprint. In our paper, we proposed a quantum algorithm that uses cosine similarity, which is one of the popular approaches usually used to mitigate device heterogeneity effects \cite{cos_sim1, cos_sim2}.

We start the section by explaining how to construct the device-independent fingerprint and how to obtain the quantum fingerprint through encoding the offline RSS vectors from the device-independent fingerprint and the online RSS vector in qubits (i.e. state preparation phase). Then, we present the details of the quantum fingerprinting matching algorithm. Finally, we give an example on how the algorithm can work in detail.

\subsection{Device-independent Fingerprint Transformation}
In this section, we describe two different techniques to handle the devices heterogeneity problem: the power ratio and the power difference. The basic idea is to transform the traditional RSS fingerprint to a device-independent fingerprint. 

\subsubsection{Power ratio}
This technique assumes that the ratio between the RSS values coming from different cell towers remains the same on the different phones. Hence, it transforms the RSS values (i.e. power) to a relative power. That is, instead of using the raw RSS from individual RPs in the fingerprint building phase (offline phase) and the raw RSS in the online matching, it uses the ratio of the RP RSS readings from \textit{\textbf{each pair}} of RPs. 

More formally, given a RSS features vector $X$= $(f_1,f_2,..,f_N)$ from $N$ RPs in the environment, where $f_i$ is the RSS coming from RP $i$, this technique transforms $X$ to $X_{r}$= $(r_{1,2},r_{1,3},..,r_{N-1,N})$, where $r_{i,j} = \frac{f_i}{f_j}$ and $\vert X_{r} \vert ={N \choose 2}$.
%


\subsubsection{Power difference}
This approach is similar to the previous approach but takes the power difference instead of the power ratio. It assumes that the difference between the RSS values coming from different RPs remains the same on the different phones.

More formally, given a RSS features vector $X$= $(f_1,f_2,..,f_N)$ from $N$ cell-towers in the environment, where $f_i$ is the RSS coming from cell tower $i$, this technique transforms $X$ to $X_{d}$= $(d_{1,2},d_{1,3},..,d_{N-1,N})$, where $d_{i,j} = f_i - f_j$ and $\vert X_{d} \vert = {N \choose 2}$.

In the next subsection, we show how to map these relative \textit{classical} feature vectors into a \textit{quantum} register.

\subsection{State Preparation}

The state preparation step aims to encode the relative classical RSS vectors, coming from the power ratio/difference,  in quantum registers. Next, the quantum fingerprinting matching algorithm is applied between the online RSS register and the offline ones.
Assume there are $N$ RPs in the environment. Therefore, the ${N \choose 2}$-dimensional normalized RSS vector from the $N$ RPs $v= (\beta_0, \beta_1, ..., \beta_{{N \choose 2}-1}), \sum_{i=0}^{{N \choose 2}-1}\beta_i^2 =1$, can be encoded using a quantum register $\ket{\delta}$ with $n = log({N \choose 2})$ qubits. Where $\ket{\delta} = \sum_{i=0}^{{N \choose 2}-1} \alpha_i \ket{i}$, and the basis state $\ket{i}$ represents the binary encoding of integer $i$ \cite{qmemory,state_prep}. Note that the ${N \choose 2}$-dimensional  RSS vector can be encoded in $log({N \choose 2})$ qubits, which is an \textit{exponential saving in the space}.

To perform the state preparation, we used the quantum random access memory (QRAM) which is a popular technique in which the binary code of the RSS measurements vector (i.e. $\beta_i$) is loaded into a qubit register in parallel and conditional rotations are performed to the qubits in the register in order to encode the RSS measurements as amplitudes in the quantum register~\cite{qmemory,rebentrost2014quantum,zhao2019quantum,state_prep}.

We give an example of how to prepare this state from classical vectors later in this section.

\addtolength{\topmargin}{+0.1cm}
\begin{figure}[h!]
	\centering
	{\includegraphics[width=0.4\textwidth]{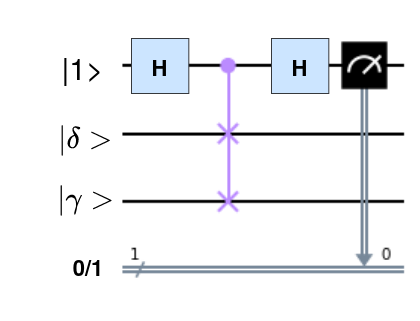}}
	\caption{Quantum fingerprint matching circuit between a single fingerprint RSS vector (encoded in qubit $\gamma$) and the online RSS vector (encoded in qubit $\delta$).}
	\label{fig:full_swap}
\end{figure}

\addtolength{\topmargin}{+0.1cm}
\begin{figure*}[t!]
	\centering
	{\includegraphics[width=0.9\textwidth]{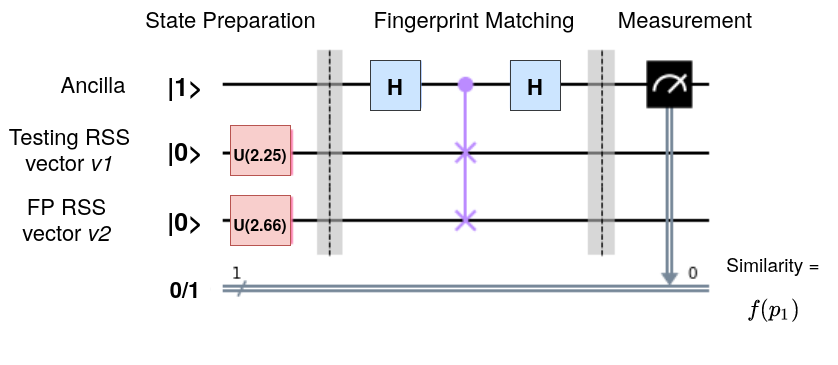}}
		\caption{A detailed example of the quantum fingerprint matching circuit using two-values RSS vectors. The circuit shows the state preparation step, i.e. how to map the testing RSS vector $(0.43,0.9)$ and training RSS vector $(0.24, 0.97)$ to a quantum state, starting from $\ket{0}$.}
		\label{fig:swap2}
	\end{figure*}

\subsection{The Quantum Cosine Similarity Algorithm}
Figure~\ref{fig:full_swap} shows the quantum circuit for calculating the cosine similarity between two normalized RSS vectors coming from applying the RSS power ratio/difference encoded in the quantum registers $\ket{\delta}$ (e.g. a test RSS vector during the online location tracking phase) and $\ket{\gamma}$ (e.g. a single fingerprint RSS vector) based on the CSWAP gate \cite{buhrman2001quantum, shokry2021quantum, shokry2021towards}. In particular, the circuit calculates:
\begin{equation}
	\textrm{sim}(\ket{\delta}, \ket{\gamma}) = \cos[2](\delta,\gamma) = |\bra{\delta}\ket{\gamma}|^2
\end{equation}

where $\cos(\delta,\gamma)$ is cosine the angle between the  two normalized vectors $\delta$ and $\gamma$.
The circuit in Figure~\ref{fig:full_swap} takes its input which is the ancilla qubit at state $\ket{1}$ and the two quantum RSS vectors encoding $\ket{\delta}$ and $\ket{\gamma}$. Then, it applies a series of gates to transform the input to the following joint state,

\begin{multline}
\frac{1}{2} \left( \sqrt{2-2 |\bra{\delta}\ket{\gamma}|^2} \ket{0}\frac{\ket{\delta}\ket{\gamma}- \ket{\gamma}\ket{\delta}}{\sqrt{2-2 |\bra{\delta}\ket{\gamma}|^2}}\right.  \\ 
+ \left. \sqrt{2+2|\bra{\delta}\ket{\gamma}|^2}\ket{1}\frac{\ket{\delta}\ket{\gamma}+ \ket{\gamma}\ket{\delta}}{\sqrt{2+2 |\bra{\delta}\ket{\gamma}|^2}} \right)
\end{multline}


Finally, the probability of measuring the top (ancilla) qubit to be 1 is $\frac{1}{2} ( 1+ |\bra{\delta}\ket{\gamma}|^2)$, which is a function of the required similarity measure between the two vectors. 
We repeat this circuit $K$ times to estimate the cosine similarity as $2 \times \#\ket{1}/K - 1$. 
Algorithm~\ref{algo1} explains our device-independent quantum fingerprinting matching algorithm. The $H$-gate is applied to the ancilla qubit in order to put it in a superposition state. The CSWAP gate is applied to entangle the ancilla qubit with the training and testing quantum registers. Finally, the $H$-gate is applied to generate the desired computation where the probability of receiving $\ket{1}$ for the ancilla qubit is a function of the required similarity.


\subsection{Example}
In this section, we illustrate the quantum fingerprint matching algorithm described in the previous section using a simple example with two-values RSS vector.  The two normalized RSS vectors to be matched are $v_1=(0.43, 0.9)$ and $v_2=(0.24, 0.97)$. Figure~\ref{fig:swap2} shows the complete circuit for calculating the cosine similarity between $v_1$ and $v_2$.

The circuit starts by the state preparation stage, i.e. mapping the RSS vectors $v_1$ and $v_2$ to the quantum equivalent $\ket{\delta}=0.43 \ket{0}+ 0.9 \ket{1}$ and $\ket{\gamma}=0.24 \ket{0}+ 0.97 \ket{1}$, respectively. This is achieved by using the $U$ gate, which is represented as,
\begin{equation}
U(\theta) = \begin{bmatrix}
	\cos(\theta/2) & -\sin(\theta/2) \\
	\sin(\theta/2) & \cos(\theta/2) 
\end{bmatrix}
\end{equation}

Where $\theta$ is double the angle between $\ket{0}$ and the quantum representation of the normalized RSS vector $\ket{v}$.

The fingerprint matching part is the same as the one described in Algorithm~\ref{algo1}. Since we have only two values in the RSS vectors, the quantum registers contain only one qubit. For the given example, the probability of measuring the first qubit to be in state 1 is 0.9765 and hence the similarity score is 0.9529. 

Note that the proposed quantum circuit needs to be repeated for each of the fingerprint locations to determine its similarity score to the test RSS vector. The fingerprint location with the highest score becomes the estimated user location.
 
\addtolength{\topmargin}{+0.1cm}
\begin{algorithm}[!t]
	\caption{\sys{} Fingerprint Matching} 
	\label{algo1} 
	\begin{algorithmic}[1]
		\Require
		\Statex 1- Two n-qubits quantum registers  $\ket{\delta}$  and $\ket{\gamma}$, storing RSS vectors of the FP and test locations, coming from the power ratio/difference techniques, to be compared. $n = log({N \choose 2})$, where $N$ is the number of RPs.
		
		\Statex 2- An ancilla qubit $\ket{a} = \ket{1}$
		
		\Statex 3- Number of iterations $K$.
		
		\Ensure
		\Statex Compute an estimate of the similarity between $\ket{\delta}$  and $\ket{\gamma}$ as $|\bra{\delta}\ket{\gamma}|^2$
		\newline
		
		\For{$k \gets 1$ to $K$}                    
		\State Apply $H(\ket{a})$
		\For{$i \gets 1$ to $n$}                    
		\State Apply CSWAP($\ket{a}$,$\ket{\delta_i}$,$\ket{\gamma_i}$)
		\EndFor
		\State Apply $H(\ket{a})$
		\State $\eta_k$ $\gets$ measurement of $\ket{a}$. 
		\EndFor
		\State return  $\frac{2}{K} \sum_{k=1}^{K} \eta_k - 1$ 
	\end{algorithmic}
\end{algorithm}

\section{Implementation and Evaluation}
\label{sec:eval}
In this section, we implement the proposed quantum localization algorithm and evaluate its performance in a real testbed. 
\subsection{Experiment setup}
\begin{figure}[!t]
	\centerline
	{\includegraphics[width=0.5\textwidth]{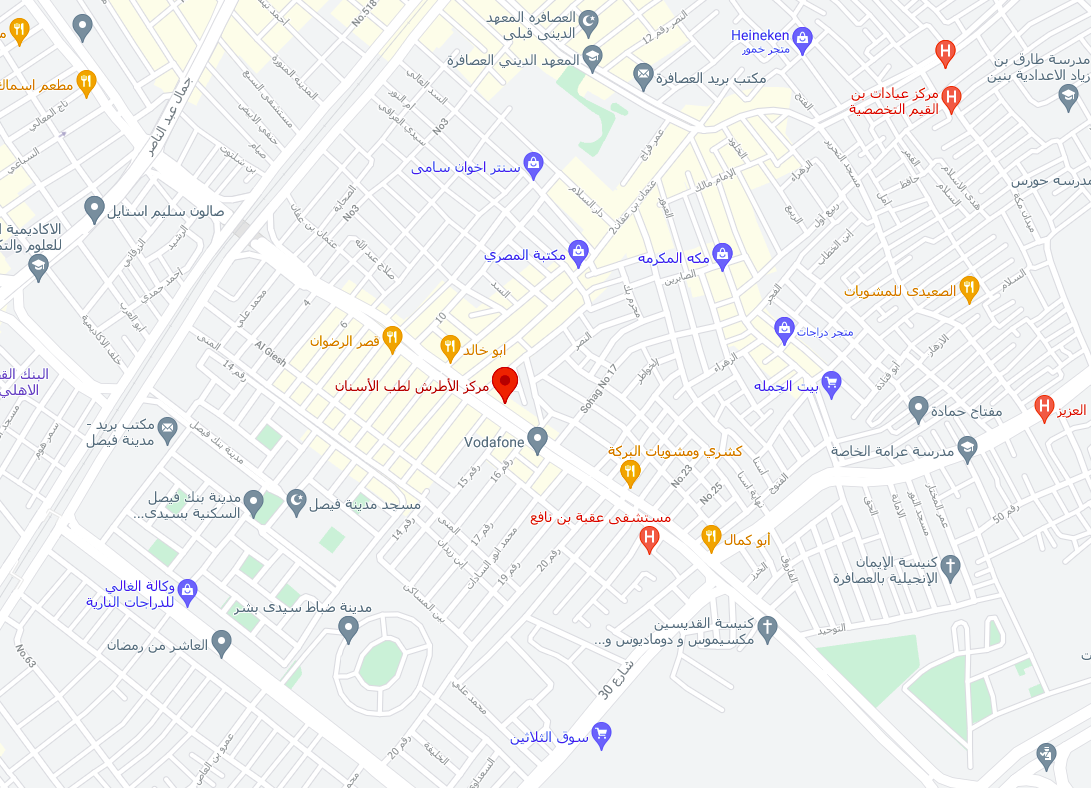}}
	\caption{The outdoor testbed.}
	\label{fig:testbed2}
\end{figure}

Figure~\ref{fig:testbed2} shows our real cellular testbed that spans an 0.2Km$^2$ outdoor urban area. The area is covered by 8 different cell-towers. Data is collected by war-driving uniformly over the entire area of interest. We also collected an independent test set. Both the fingerprint and test locations are uniformly distributed over the entire area of interest. 
We use different Android devices for data collection including HTC Nexus One, Prestigio Multipad Wize 3037 3G, HTC One X9 and Motorola Moto G5 plus phones among others. The deployed collector software collects GPS ground-truth locations, received signal strengths, and timestamps.

The 8 cell-towers RSS can be encoded in 5 qubits after applying RSS power ratio or difference techniques. Therefore, we require a total of 11 qubits for running the circuit in Figure~\ref{fig:full_swap}, 10 qubits for encoding the training and testing RSS vectors and one for the Ancilla qubit. We use the IBM quantum machine simulator, which supports up to 28 qubits.

%


%



\subsection{Comparison Between the Different Heterogeneity Handling Techniques}
\begin{figure}[!t]
	\centerline
	{\includegraphics[width=0.5\textwidth]{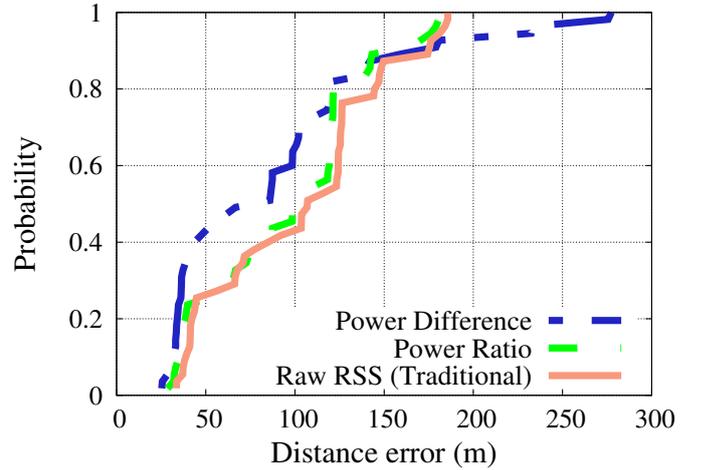}}
	\caption{Effect of device heterogeneity techniques.}
	\label{fig:cdf_rd_8}
\end{figure}

Figure~\ref{fig:cdf_rd_8} and Table~\ref{comparison} show a comparison between the traditional raw RSS, the quantum power ratio, and the quantum power difference fingerprinting techniques. Specifically, the results show that \sys{}, through its heterogeneity handling techniques, can achieve a better localization accuracy than traditional raw RSS localization by more than 20\% in the median error.  The results also show that the power difference technique is superior to the power ratio technique. This can be explained by noting that the RSS reported by the phone API is in log-scale.

\begin{table}[!t]
	\centering 
	\caption{Comparison between quantum and classical localization systems error quantiles. Numbers between parenthesis indicate percentage of enhancement over raw RSS.}
	\label{comparison}
	\begin{tabular}{|p{4cm}|p{1cm}|p{1cm}|p{1cm}|}\hline
		System & Q1 & Median & Q3\\ \hline \hline	
		Raw RSS - Traditional & 40  & 103  & 121 \\ \hline	
		\sys{} - Power Ratio & 39 \textbf{(2.6\%)}&  98 \textbf{(5.1\%)}& 121 \textbf{(0\%)}\\ \hline
		\sys{} - Power Difference  & 36 \textbf{(11.1\%)}  &  85.5 \textbf{(20.5\%)} & 119 \textbf{(1.7\%)}\\ \hline 
	\end{tabular}
\end{table}





\subsection{Comparison with Classical Localization}

\begin{figure}[!t]
	\centerline
	{\includegraphics[width=0.5\textwidth]{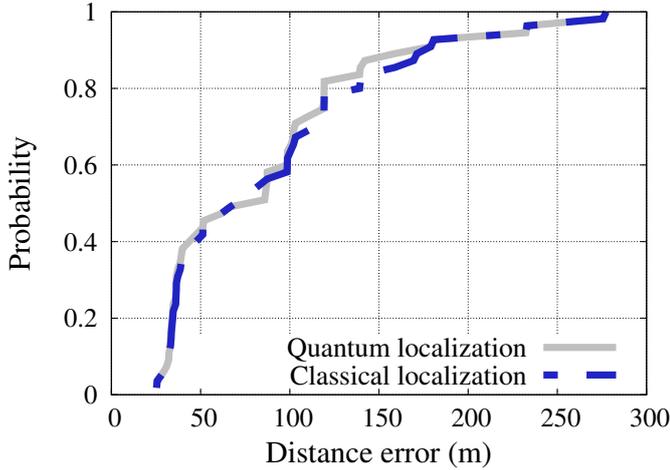}}
	\caption{CDF of localization error for quantum and classical localization using the power difference technique.}
	\label{fig:cdf_d}
\end{figure}

Figure~\ref{fig:cdf_d} shows the CDF of distance error for the quantum and classical localization system using the same device-independent features (i.e. after applying the power difference heterogeneity technique). The figure shows that \sys{} has the same accuracy as the classical one,  but with the potential exponential saving in time and space.

\subsection{Effect of Number of Shots}

\begin{figure}[!t]
	\centering
	{\includegraphics[width=0.5\textwidth]{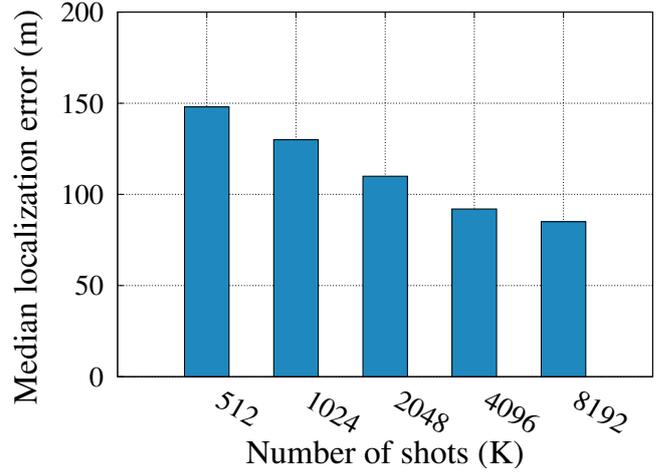}}
	\caption{Effect of number of shots on the quantum power difference localization error.}
	\label{fig:dshots}
\end{figure}

Figure~\ref{fig:dshots} shows the impact of increasing the number of algorithm iterations, i.e. re-running the system (parameter $K$ in Algorithm~\ref{algo1}), on the power difference quantum localization accuracy. It is evident from the figure that increasing the number of iterations leads to a better localization accuracy for both techniques till it starts saturating around 4096 shots.

\subsection{Discussion}
\label{sec:discuss}


Last section shows that the device-independent fingerprinting localization techniques have accuracy that is better than the traditional raw RSS one in the real testbed used in our experiments.

\begin{figure}[t!]
	\centerline
	{\includegraphics[width=0.5\textwidth]{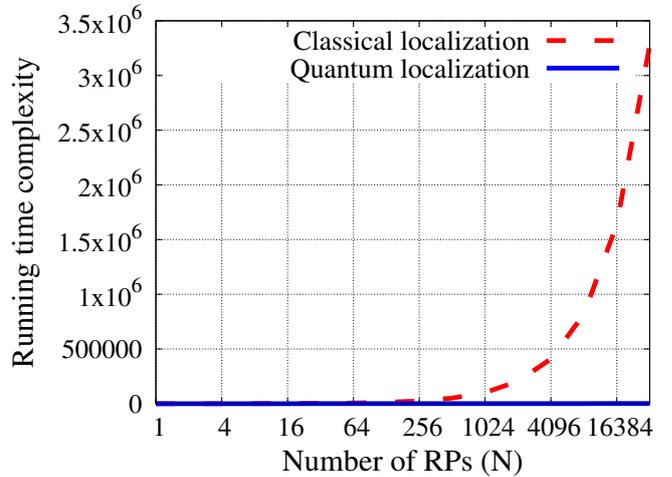}}
	\caption{Running time complexity for the quantum and the classical algorithms.}
	\label{fig:complexity}
\end{figure}

The device-independent classical version of the localization system requires $o(N^2M)$ space and matching runs in $o(N^2 M)$, for $N$ RPs and $M$ fingerprint locations. On the other hand, \sys{} requires $o(M \log N)$ for both space and time. This is an exponential enhancement in both space and running time.
Figure \ref{fig:complexity} explains how the running time complexity for the quantum and the classical localization systems scale with the increase of the number of RPs for a fixed number of locations in the fingerprint. The figure confirms the \textbf{\textit{exponential saving in running time}} of the proposed quantum algorithm.
This is important in many scenarios such as worldwide localization and IoT applications, where there are alot of RPs with different technologies (cellular, WiFi, BT, etc). Moreover, in order to get a better accuracy, one needs to fuze the signal coming from the significant number of RPs in the environment. Hence, the exponential saving in time and space becomes a must.

\sys{} is not only useful for the online fingerprint matching process but also for offline state preparation (quantum fingerprint construction) as it can reduce the size of the fingerprint RSS vectors from $o(N^2)$ to o(log(N)). This is important not only to speed up the matching process but also to save the space required for downloading and storing the large-scale worldwide fingerprint from the server to mobile phones if the fingerprint matching is performed on the mobile phones. In addition, quantum co-processors (similar to GPUs) have started  to appear~\cite{frangou2019first} which can enable quantum localization on mobile devices.  

Finally, the space and time complexity of \sys{} can be reduced to $o(\log(N M))$ by encoding all the fingerprint locations data in one circuit.  This might have a significant implications to several localization and spatial algorithms.

\section{Conclusion}
\label{sec:conclude}

In this paper, we have presented, \sys{}, a practical device-independent quantum fingerprinting algorithm. \sys{} can provide high accuracy localization with different phone types while leading to an exponential saving in the space and the running time of the current fingerprinting localization systems. Results from deploying \sys{} in a real testbed confirm its advantages compared to the traditional classical techniques. Moreover, the proposed device-independent features achieves a median distance error of 85.5m, which is 20\% better than the traditional fingerprinting methods. 

Currently, we are exploring different quantum similarity metrics, obtaining theoretical quantum bounds on performance, and optimizing other spatial algorithms using quantum computing.

\balance
\bibliographystyle{plain}
\bibliography{main.bib}

\end{document}